\documentclass[11pt,twoside]{article}

\usepackage{galev06}
\usepackage{epsf}
\usepackage{psfig}
\usepackage{lscape}

\begin{document}
\title{The buildup of stellar mass of galaxies 
in clusters during the last two thirds of the universe age and 
associated methodological issues}   
\author{S. Andreon}   
\affil{INAF-Osservatorio Astronomico di Brera, Milano, Italy }    

\begin{abstract} 

We have measured the 3.6 $\mu$m luminosity and mass evolution of about
1000 galaxies in 32 clusters at $0.2<z<1.25$ with a special
attention to methodological issues, as emphasized in this proceeding
contribution.
We find that the luminosity of our galaxies evolves as an old and
passively evolving population formed at high redshift without any need
for additional redshift-dependent evolution. Models with a prolonged
stellar mass growth are rejected by the data with high confidence.  The
data also reject models in which the age of the stars is the same at all
redshifts.  Similarly, the characteristic stellar mass evolves, in the
last two thirds of the universe age, as expected for a stellar
population formed at high redshift. Together with the old age of stellar
populations derived from fundamental plane studies, our data seems to
suggest that massive early-type cluster galaxies have been completely
assembled at high redshift, and not only that their stars are old. The
quality of the data allows us to derive the LF and mass evolution 
homogeneously over the whole redshift range, using a single estimator.
The Schechter function describes the galaxy luminosity function well.
The characteristic luminosity at $z=0.5$ is found to be 16.30 mag,
with an uncertainty of 10 per cent. 

\end{abstract}


\section{Introduction}   

The luminosity function (LF) is the basic statistic  used to understand
galaxy properties, giving  the relative frequency of galaxies of a given
luminosity in a given volume.
Since starlight at 3.6 $\mu$m very nearly follows the Rayleigh-Jeans limit
of blackbody emission for $T > 2000$ K, the colors of both early- and
late-type stars are similar.  There is virtually no dust extinction at this
wavelength either, since any standard extinction law predicts only a few
percent of the extinction of optical wavelengths. The 3.6 $\mu$m
light therefore traces the stellar mass distribution free of dust
obscuration effects (Pahre et al. 2004). Thus, a useful approach to
understanding how galaxies form is to track their growing stellar mass,
measured through the evolution of the 3.6 $\mu$m LF.

Several studies found LF results consistent with the behavior of a
simple, passive luminosity evolution model in which galaxies form all their
stars at high redshift and thereafter passively evolve. However,
seldom these studies address which models 
data exclude. What it actually interesting is whether some 
plausible scenario can be rejected by the data, because sometime
a (wrong) scenario cannot be rejected simply because of the 
low information content of the data in hand. Furthermore, 
previous works
rarely address the topic of sample representativity.
Specifically, clusters in the sample 
are typical or atypical clusters? All types of
clusters are included in the studied sample? Some type of
cluster is over- or under- represented in the sample?
The above questions are closely related to the following one:
can a result found to hold on a sample of unknown representativity
(the studied sample of clusters)
be generalized to the parent distribution (i.e. to clusters)?
Results obtained on uncontrolled sample rarely hold in
general (and when they hold, they hold by good chance).

Throughout this paper we assume $\Omega_M=0.3$, $\Omega_\Lambda=0.7$ 
and $H_0=70$ km s$^{-1}$ Mpc$^{-1}$. 

\section{Data \& data reduction}

IR data were obtained with the IRAC (Fazio et al. 2004) on the Spitzer
Space Telescope (Werner et al. 2004). Optical data come from the wide
field imager MOSAIC-II at the 4m CTIO. Details on data (and more)
can be found in Andreon (2006).

\section{The sample}

The cluster sample studied in this paper consists of 32 colour--selected
clusters, all spectroscopically confirmed. The availability of
spectroscopic redshifts is an essential difference from some previous
works having photometric
redshift only.  
In works lacking a cluster spectroscopic confirmation,  errors on LF
parameters (and evolution) are unduly kept small by ignoring the
uncertainty due to the contamination in the cluster sample by other,
often only apparent, structures (e.g. line of sight superpositions).
As shown in Yamada et al. (2005) with a few examples, this occurs
frequently to works lacking a cluster spectroscopic confirmation.
Working with spectroscopically
confirmed clusters, we are not affected by the uncertainty due to
unrelated structures entering in the sample. Furthermore, we are not
affected by uncertainties on cluster distances that instead affect works
using photometric redshifts.

The clusters were detected as spatially localized galaxy overdensities of
similar optical colour, as described in Andreon et al. (2003;
2004a,b). The studied cluster sample is not a volume complete sample, 
nevertheless it densely samples the explored Universe volume, up to
$z\sim1$, that should make the studied sample representative of
typical clusters, and our results of high generalization power.

Most of our clusters are at the bottom of the Abell richness scale and are
not rich systems. This is unsurprisingly, because the mass (richness)
function of clusters is steep. Instead, other samples studied in literature
consider rich (and therefore rare) clusters, biasing the representativity of
them with respect to clusters of modest richness and larger abundance.

The 32 clusters are distributed in redshift as shown in Fig 1. In
particular, 6 clusters have $z>0.99$ and 8 clusters have $z>0.85$.
These clusters are essential to discriminate among different histories 
of star formation and mass assembly, because over a short redshift
range different histories do not differ too much 
(see right panel of Fig 2). To our best knowledge, no previous
work scores better than our work in the number of clusters at 
$z>0.99$ or $z>0.85$.

\begin{figure}
\centerline{\psfig{figure=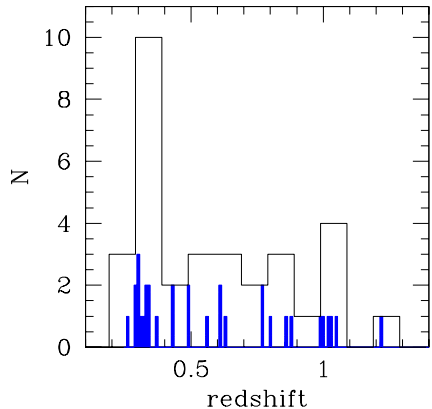,width=6truecm}\quad
\vbox{\hsize=7 truecm \caption[h]{Redshift distribution of the studied clusters. Solid/open histograms mark
0.01/0.1 bin width in redshift.}}}
\end{figure}

\section{The method: LF determination}

The determination of the LF uses state of the art 
statistical methods: an
improved likelihood function (Andreon, Punzi \& Grado 
2004) and Bayesian methods for model selection.
Incongrounces, logical contradictions, circular arguments,
arbitrary bins, unphysical values for quantities, all
have been avoided by a rigorous statistical analysis (see Andreon et al.
2006 and Andreon 2006). For example, in order to determine
$M^*$ our statistical analysis
don't oblige us to keep fixed
the nuisance parameter $\alpha$, because unconstrained, 
as instead other
analysis oblige.
If $\alpha$ is unconstrained, 
what statistical theorem
allows other works to keep it to be a fixed (and arbitrary) 
value? By the way, keeping nuisance parameters 
fixed contradicts the sum rule of probabilities
and underestimates the uncertainty on $M^*$ (the latter as 
re-discovered by Andreon 2004). Furtheremore, other works
bin clusters in redshift bins,  and forget to 
investigate whether the found/missed effect
is related to some binning choices (bin size, bin location,
sub-optimal binning strategy, etc.).  Instead, we don't bin
clusters in redshift bins and therefore we are not affected
by bin resolution/choice.
The analysis also account for the younger age of the universe
(and therefore of stars) at hight redshift. 
Full details are described in Andreon (2006), where 
a traditional analysis is also presented for old-fashioned
readers. 

Grasil (Silva et al. 1998) models are used to convert light in 
stellar mass and to convert apparent to absolute luminosities.

\begin{figure*}
\centerline{\psfig{figure=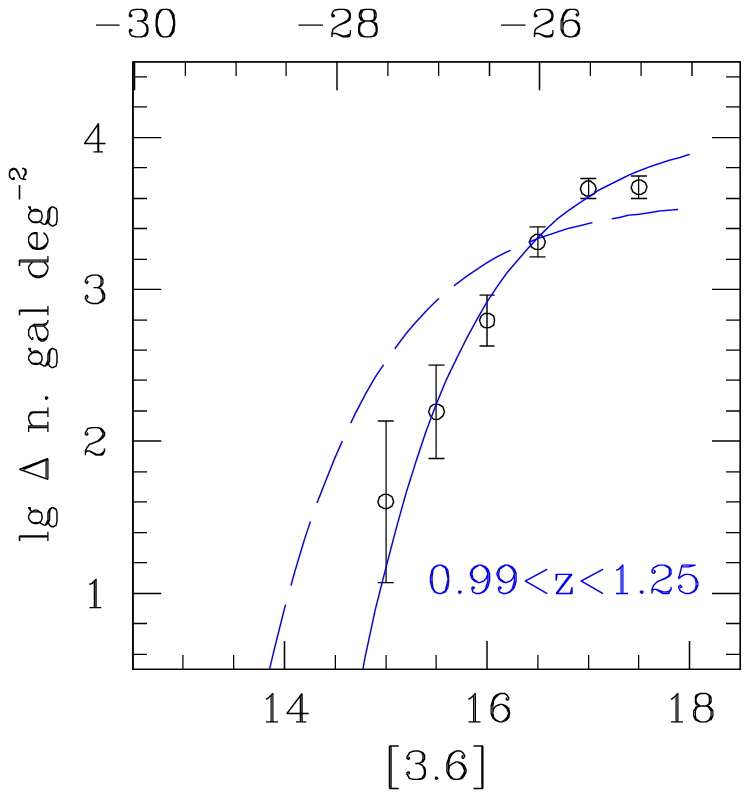,width=6truecm}%
\psfig{figure=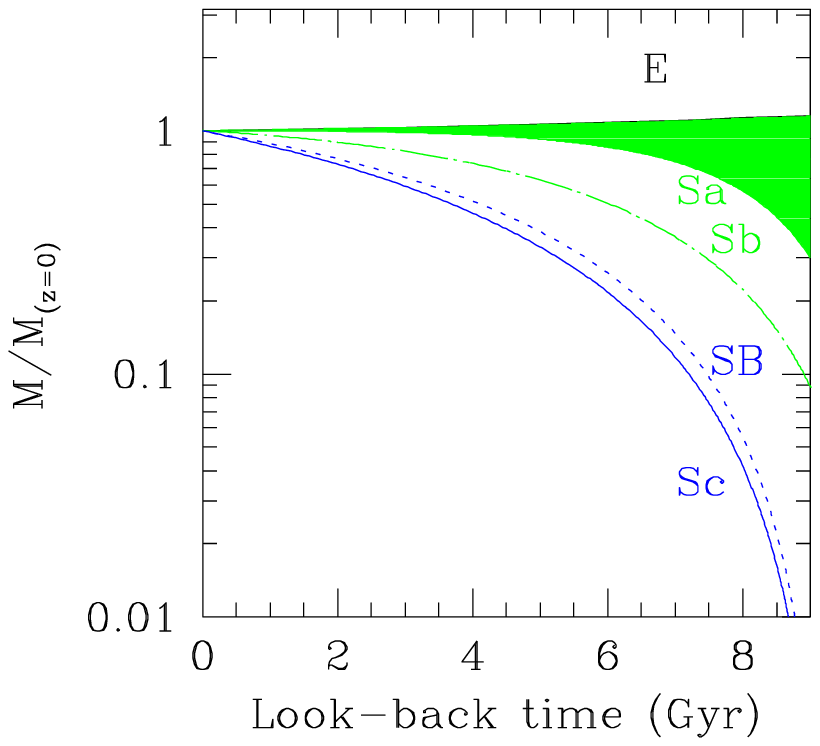,width=6truecm}}
\caption[h]{
{\it Left:} Composite LF in the $[3.6]$ band as a function of
apparent (lower abscissa) and absolute (upper abscissa) magnitudes, 
computed using standard methods.  
The solid and long dashed
curves are fit to unbinned counts. The solid curve refers to a fit with
$m^*$ free, whereas the dotted curve is a fit with $m^*$ held fixed to the
value observed at $0.25< z <0.40$. {\it Right:}
Stellar mass evolution for several stellar mass growth models 
(curves) and as derived from our data (shaded area), using
the full statistical analysis.  
The region allowed by the data is part of the shaded green area.
A factor two growth in mass is rejected by the data. Details in
Andreon (2006)}
\end{figure*}


\section{Results}

Astronomical results are summarized in the abstract section. 
Shortly, only a model is viable for the evolution of cluster galaxies: 
a model in which there is almost no mass growth during
the last two third of the universe age. All the remaining considered
models are rejected by the data.
Figure 2 is a pictorial excerpt of the results. 
A full report, including methodological details, 
is presented in Andreon (2006).

\acknowledgements 
We acknowledge financial contribution from contract ASI-INAF I/023/05/0 


\end{document}